\documentclass[conference]{IEEEtran}

\usepackage{graphicx}
\usepackage{epsfig}
\usepackage{subfig}
\usepackage{epstopdf}

\usepackage{floatrow}
\begin{document}

%
\title{Providing Physical Layer Security for Mission Critical Machine Type Communication}

\author{{Andreas Weinand, Abhijit Ambekar*, Michael Karrenbauer, and Hans D. Schotten} \\
	Chair for Wireless Communication and Navigation\\
	University of Kaiserslautern, Germany\\
	**German Research Center for Artificial Intelligence (DFKI), Kaiserslautern, Germany\\
	Email: \{weinand, karrenbauer, schotten\}@eit.uni-kl.de, abhijit.ambekar@dfki.de
}


%


\maketitle

\begin{abstract}
The design of wireless systems for Mission Critical Machine Type Communication (MC-MTC) is currently a hot research topic. Wireless systems are considered to provide numerous advantages over wired systems in industrial applications for example. However, due to the broadcast nature of the wireless channel, such systems are prone to a wide range of cyber attacks. These range from passive eavesdropping attacks to active attacks like data manipulation or masquerade attacks. Therefore it is necessary to provide reliable and efficient security mechanisms. One of the most important security issue in such a system is to ensure integrity as well as authenticity of exchanged messages over the air between communicating devices in order to prohibit active attacks.
In the present work, an approach on how to achieve this goal in MC-MTC systems based on Physical Layer Security (PHYSEC), especially a new method based on keeping track of channel variations, will be presented and a proof-of-concept evaluation is given. 
\end{abstract}

%
\IEEEpeerreviewmaketitle

\section{Introduction}
\label{attacker_model_section}
{\let\thefootnote\relax\footnote{This is a preprint, the full paper has been published in Proceedings of IEEE 21th Conference on Emerging Technologies \& Factory Automation (ETFA 2016), \copyright 2016 IEEE. Personal use of this material is permitted. However, permission to use this material for any other purposes must be obtained from the IEEE by sending a request to pubs-permissions@ieee.org.}}
\IEEEpubidadjcol

\begin{figure}[h]
\centering
\includegraphics[width=3.5in]{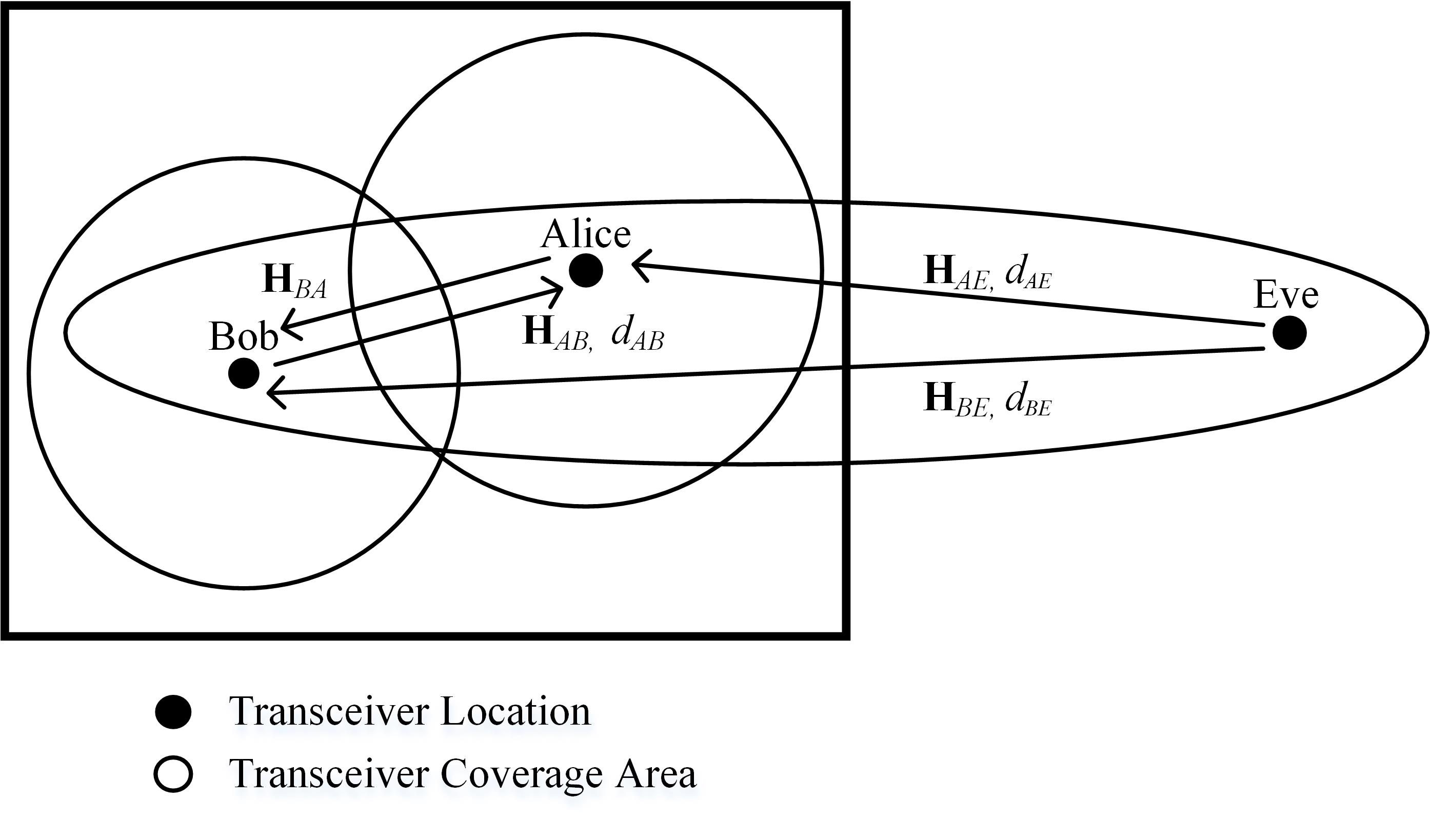}
\caption{System and Attacker Model}
\label{attacker_model}
\end{figure}

\label{intro}
Wireless communication systems have a lot of advantages compared to wired communication systems, especially in industrial environments. These are, for example, higher flexibility, lower cost in manufacturing and maintenance as well as the enabling of new applications in the context of the ongoing fourth industrial revolution. On the other hand the wireless nature of the underlying channel provides a huge potential for diverse attacks. Every possible participant inside the coverage of the system is able to passively and actively interact with the transmitted information and consequently able to interact with the served applications as well. In this work, a promising approach on how to secure such a system in an efficient way based on PHYSEC is presented.\\
Regarding the type of communication considered in this work, which is MC-MTC, closed loop control applications might be a possible use case. For this instance, primarily the safe and secure operation of the controlled machines or processes and secondly the nondisclosure of intellectual property such as process control parameters, machine configuration data or even simple information such as the production volume of controlled processes have to be ensured. 
Due to security flaws in today commonly applied protocols and systems in industrial scenarios, it is essential for future systems to provide adequate security measures which rely on both, conventional cryptography, as well as consideration of physical conditions of the present environment.
A typical scenario for an attacker in such a system could be that he is located outside of a factory environment using advanced equipment such as directed antennas and high sensitivity receivers to maximize the range of the attacked system to his benefit (see Fig. \ref{attacker_model}), as well as appropriate computational power and perfect knowledge of the underlying communication protocol to accomplish passive and active attacks. In the case of passive attacks, eavesdropping and traffic analysis are considered and for the case of active attacks, replay attacks and masquerade attacks such as man-in-the-middle attacks or address spoofing are considered. It is not assumed that the attacker is gaining physical access to the environment of the deployed system to accomplish invasive attacks such as hardware modification. Further, Denial-of-Service attacks due to jamming are not considered as well. It is assumed, that legal communicating participants have already carried out initial user authentication to each other and have set up initial trust in a secure way.\\

In order to prevent and detect the mentioned scenarios of possible attacks, suitable security measures have to be applied. Even though this can be provided by means of conventional security techniques, these techniques also bring some disadvantages with them, for example the additional overhead introduced by message integrity check codes to ensure message integrity and authenticity, which is also the goal of this work. Therefore, we propose an approach based on PHYSEC in our work, which ensures this in an efficient way based on so called Channel Profile Monitoring (CPM). The main idea is to exploit characteristics of the wireless channel in time and spatial domain. As mentioned, MC-MTC and closed loop control applications are considered here, which in combination seems to be a perfect case for our approach, as we can assume that frequent and periodic data transmissions and with this channel estimation at the same rate (e.g. once per ms) is carried out.
  

The remainder of the work is organized as follows. In section \ref{related work} we give a short overview on related work with respect to previous considered approaches and in section \ref{PHYSEC_integration} our approach on how to fulfill the above mentioned security objectives regarding active attacks based on PHYSEC is introduced. In section \ref{results} we give insight in first results of our work and section \ref{CONC} finally concludes the paper.

\section{Related Work}
\label{related work}
In recent years, several approaches on exploiting the wireless channel for security purposes, also known as PHYSEC, have been investigated. Many works have focused on extracting secret keys between two communicating devices, such as \cite{Guillaume.}, \cite{Zenger.2014}, \cite{Ambekar.2014}. The focus of our work is on guaranteeing secure transmission with respect to integrity and authenticity of data packets from one device to another, which has also been considered in a couple of recent works. For example in \cite{Xiao.2007} an approach based on channel measurements and hypothesis testing is presented for static scenarios and later in \cite{Xiao.2008} extended to time-variant scenarios. Though the results of these works are plausible, they are obtained from simulations of the wireless channel only. Therefore, in our work we will focus on real world channels. In \cite{Pei.2014} two approaches based on machine learning, Support Vector Machine and Linear Fisher Discriminant Analysis, are presented. A Gaussian Mixture Model based technique is applied in \cite{Gulati.2013}. The approach considered in \cite{Tugnait.2010} is similar to our approach, as they propose a CSI-based authentication method. However, they consider a single carrier system, while we will focus on a multi carrier system for evaluation later in section \ref{results}. The second approach considered in \cite{Tugnait.2010} is whiteness of residuals testing. In \cite{Shi.2013} an RSS-based approach for body area networks is presented. The work in \cite{Refaey.2014} considers a multilayer approach based on OFDM to guarantee authentication of TCP packets.

\section{Channel Profile Monitoring for MC-MTC}

\label{PHYSEC_integration}
In this section we describe how to protect MC-MTC from the active attacks mentioned in section \ref{attacker_model_section} by exploiting PHYSEC techniques, actually Channel Profile Monitoring based on frequent channel measurements.

\subsection{Drawbacks of MIC for MC-MTC}
In commonly deployed systems for industrial applications such as IEEE 802.15.4 based systems, integrity checking of MAC (media access control layer) payload is provided in form of message authentication codes (MAC, to not confuse with these abbreviations, in the following MIC is used to refer to message integrity checking based on message authentication codes and MAC is used to refer to media access control layer). These calculate a check sum which is added to the the actual MAC payload. In IEEE 802.15.4 based systems either a 4, 8 or 16 Byte MIC sum is added to the respective payload. This leads to an overhead for the MAC payload on data frames of at least $3.54\%$ if 4 Byte MIC is considered as shown in Tab.\ref{mic_overhead} (without any other MAC overhead such as header or footer and if maximum MAC payload size is considered). In our work we assume that the payload of a MC-MTC packet has a length of 40 Byte. The dimension of this assumption is e.g. confirmed by \cite{Osman.2015}. Then the payload overhead regarding the shortest possible MIC size of 4 Byte would be already $10\%$. Consequently the application of MIC sums seems to be not appropriate for this use case. Another important issue is that only the MAC payload of messages is secured. An attacker can e.g. still manipulate Bits in the MAC header to fake the source ID of the message or even worse unset security options.  
\begin{table}[h]
\centering
\resizebox{\textwidth}{!}{  
\begin{tabular}{lccc}
\hline
System & MIC size & MAC payload & MIC overhead\\
\hline
IEEE 802.15.4 & 4 & 113 & 3.54\% \\
IEEE 802.15.4 & 8 & 113 & 7.08\% \\
IEEE 802.15.4 & 16 & 113 & 14.16\% \\
MC-MTC & 4 & 40 & 10\%\\
MC-MTC & 8 & 40 & 20\% \\
MC-MTC & 16 & 40 & 40\% \\
\hline
\end{tabular}
}
\caption{MIC overhead for IEEE 802.15.4 and considered MC-MTC (size of MIC and MAC payload given in Byte)}
\label{mic_overhead}
\end{table}

\subsection{New Approach based on Channel Profile Monitoring}
Due to these drawbacks of MIC sum based integrity checking, a more promising approach is to monitor the channel profile by taking frequent channel measurements into account. In contrast to PHYSEC techniques such as secret key generation, which is based on the assumption that there is a lot of temporal variation in a wireless channel, our approach relies on the fact that the wireless channel does not vary significantly during subsequent channel measurements. However, the same idea that yields for both is to  make use of the advantage of the fast spatial decorrelation of wireless channels. For CPM in particular, this means, that e.g. Alice receives packets from the legal transmit node Bob and estimates the actual channel conditions $\mathbf{H}_{AB}^{(k)}$. The idea is now to compare the channel measurement which is experienced by Alice during reception of a packet from Bob at time $k$ to the channel measurement of the previous reception at time $k-1$ and calculate the mean square error (MSE)

\begin{equation}
e_{AB}^{(k)}=\mathrm{MSE}(\mathbf{H}_{AB}^{(k)},\ \mathbf{H}_{AB}^{(k-1)})
\end{equation} 

with 
\begin{equation}
\mathrm{MSE}(\mathbf{X}, \mathbf{Y})=\frac{1}{LM}\sum_{i=0}^L \sum_{j=0}^M (\mathbf{X}_{i,j} - \mathbf{Y}_{i,j})^2 
\end{equation}
of two matrices $\mathbf{X}$ and $\mathbf{Y}$ of size $L\mathrm{x}M$.
If the error $e_{AB}^{(k)}$ is now below a certain threshold $e_{th}$, Alice assumes that there has been no manipulation attack performed on that packet and that it was send by the legal transmitter node Bob yielding \mbox{$e_{AB}^{(k)} < e_{th}$}. This is based on the mentioned assumption that between two subsequent receptions from one transmitter the channel is considered as approximately constant \mbox{($\mathbf{H}_{AB}^{(k)} \approx \mathbf{H}_{AB}^{(k-1)}$)}, i.e. these two subsequent measurements lie within the coherence time of the channel.\\
If now an attacker Eve tries to manipulate Bobs transmissions or insert packets to masquerade as Bob, the channel profile measured by Alice differs at a much higher degree so that $\mathbf{H}_{AE}^{(k)} \ne \mathbf{H}_{AB}^{(k-1)}$ and consequently $e_{AE}^{(k)} \geq e_{th} > e_{AB}^{(k)}$. Due to the distance between the attacker node Eve and the legal transmitter node Bob, Eve is not able to masquerade without further effort due to the mentioned fast spatial decorrelation property of the channel. As a result of this, a receiver node will deal with a received packet as follows after calculating the error $e^{(k)}$ at time $k$:
\begin{equation}
     \mathrm{Receive\ packet} \left\{\begin{array}{ll} \mathrm{process\ it}, & if\ e^{(k)} < e_{th} \\
         \mathrm{drop\ it}, & if\ e^{(k)} \geq e_{th}\end{array}\right.
\end{equation}

\section{Initial Experimental Results}
\label{results}


In this section we give insight in first existing results of our work. To evaluate our concepts, we use USRP N210 SDR platforms from Ettus Research with SBX daughterboards. We use GNURadio OFDM transmitter and receiver blocks to process data packets and perform channel estimation on each received data packet. A setup with an FFT size of $64$ is considered and $48$ data subcarriers. The cyclic prefix length is $16$ samples at a baseband sample rate of $3.125$ MSps, whereas the carrier frequency is $2.45$ GHz. For each received data packet, the initial channel taps are calculated based on the known Schmidl and Cox preamble \cite{Schmidl.1997} which is also used to calculate the coarse frequency offset. Each known symbol is followed by $4$ data symbols yielding a time resolution of 128 $\mu$s for the channel measurements. As a first step, we consider a static setup where all participants do not move during transmitting and receiving.

\begin{figure}[h!]
\centering
\subfloat[Mean Square Error]{\includegraphics[width=\textwidth]{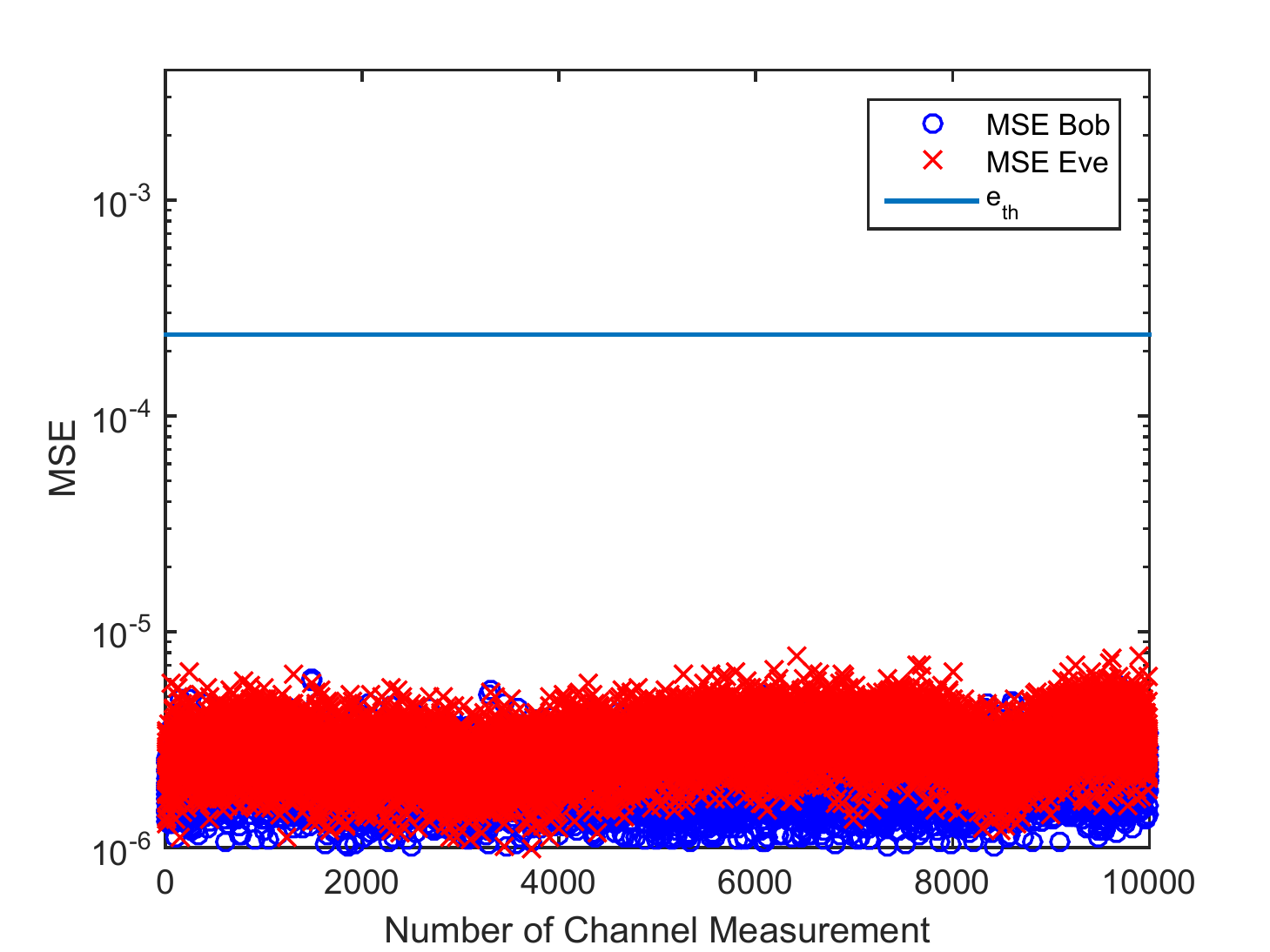}%
\label{mse_time}}
\vfil
\subfloat[Pearson Correlation Coefficient]{\includegraphics[width=\textwidth]{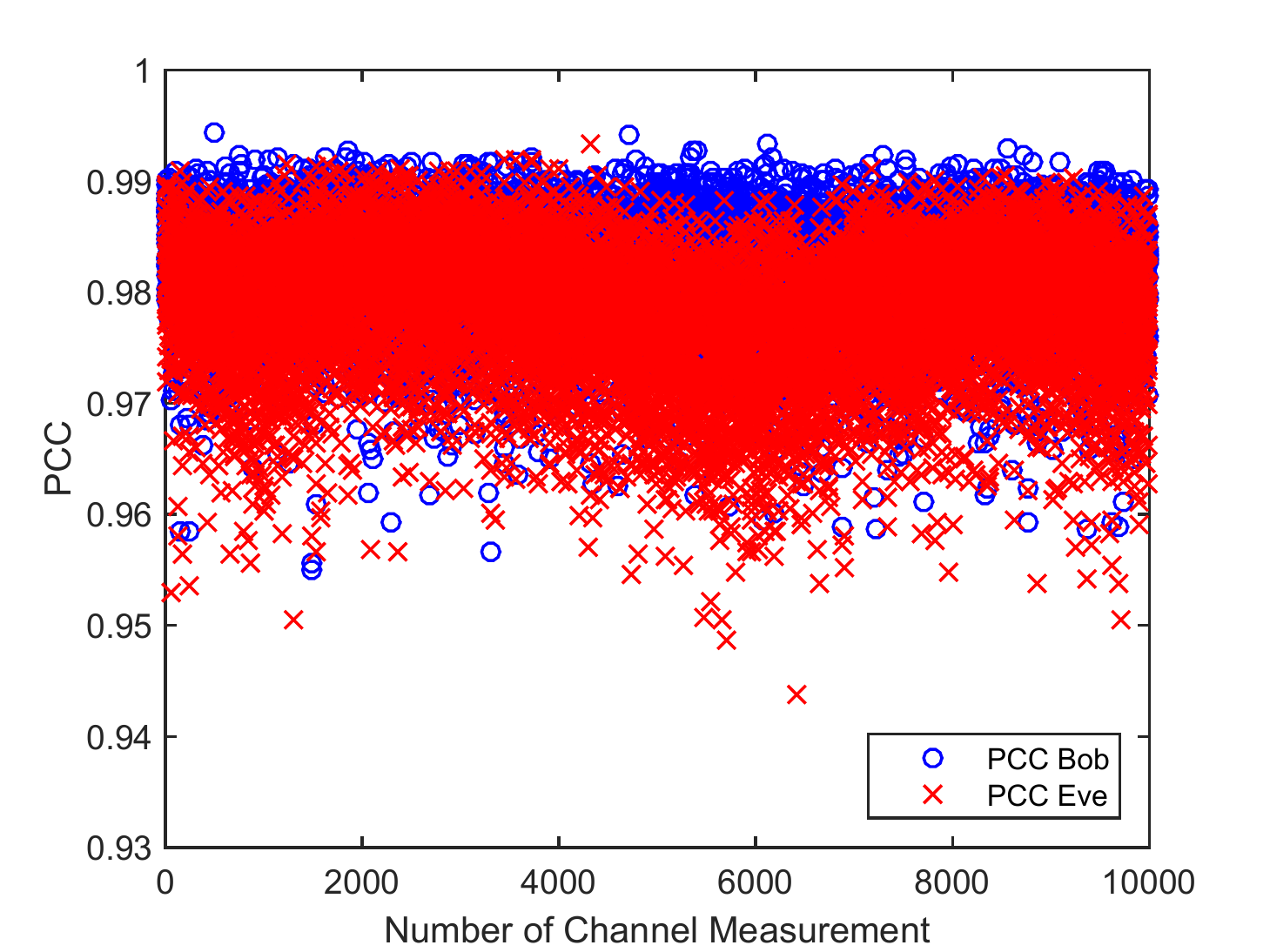}%
\label{pcc_time}}
\caption{Setup with Bob and Eve at the same location}
\label{results_time}
\end{figure}

\begin{figure}[h!]
\centering
\subfloat[Mean Square Error]{\includegraphics[width=\textwidth]{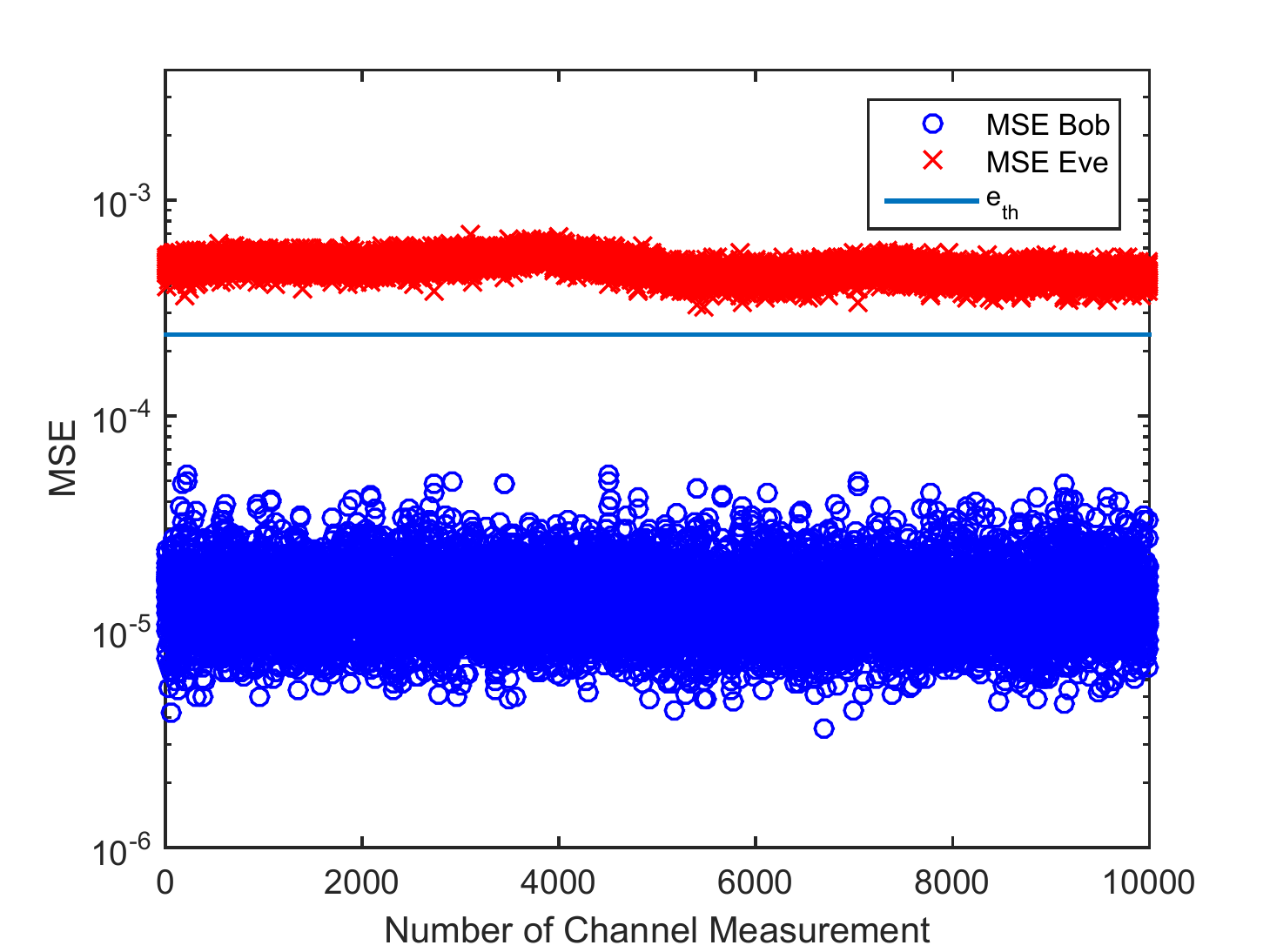}%
\label{mse_space}}
\vfil
\subfloat[Pearson Correlation Coefficient]{\includegraphics[width=\textwidth]{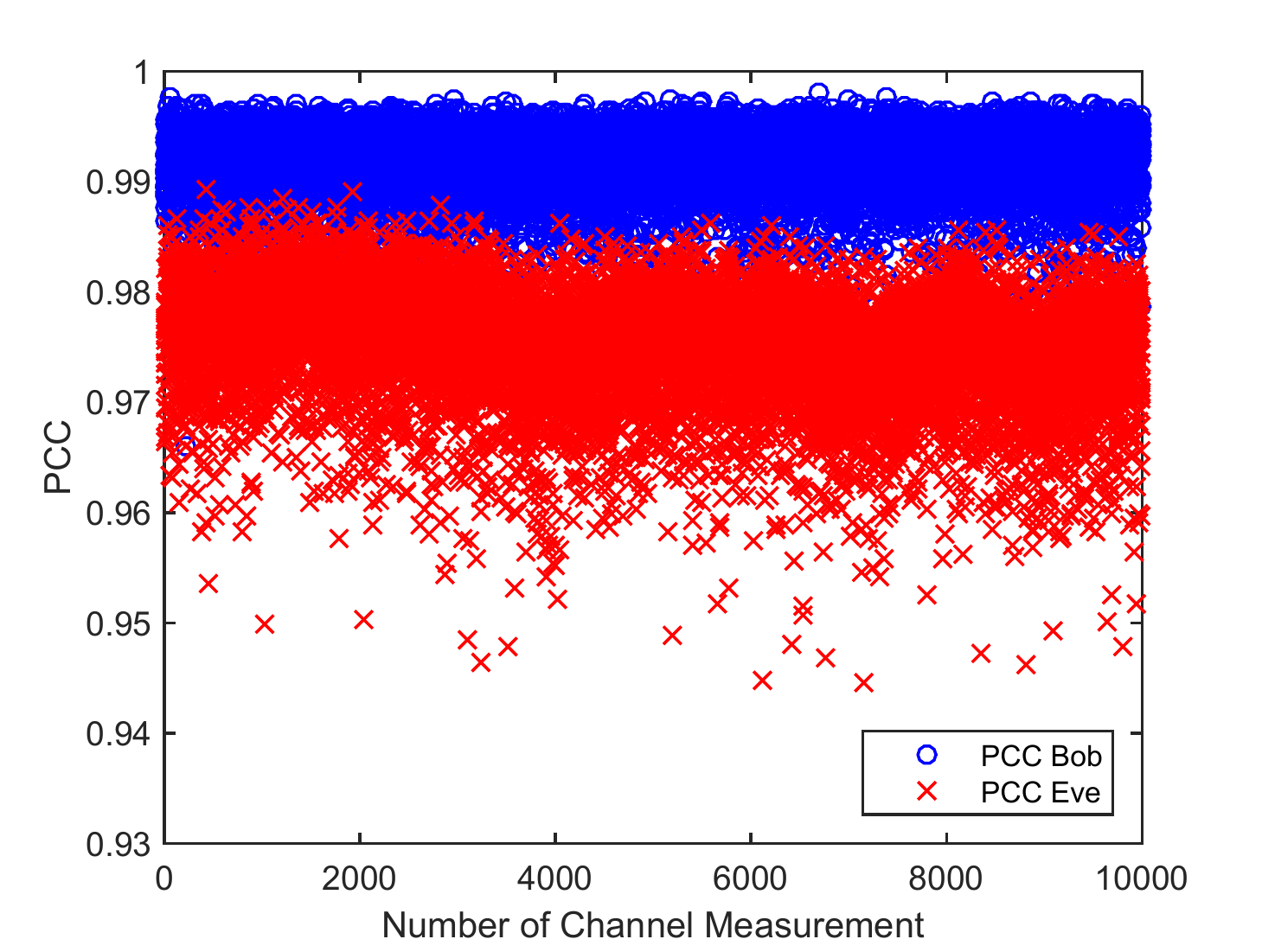}%
\label{pcc_space}}
\caption{Setup with Bob and Eve at different locations}
\label{results_space}
\end{figure}

We now calculate the elements of the error vectors $\mathbf{e}_{AB}$ and $\mathbf{e}_{AE}$ as
\begin{equation}
e_{AB}^{(k)}=\mathrm{MSE}(\mathbf{h}_{AB}^{(k)},\ \mathbf{h}_{AB}^{(k-1)})
\label{eq:}
\end{equation}
and 
\begin{equation}
e_{AE}^{(k)}=\mathrm{MSE}(\mathbf{h}_{AE}^{(k)},\ \mathbf{h}_{AB}^{(k-1)})
\label{eq:}
\end{equation}
with $\mathbf{h}_{AB}^{(k)}$ and $\mathbf{h}_{AE}^{(k)}$ being the channel taps in frequency domain of size $1\mathrm{x}M$, with \mbox{$M=64$}, derived by Alice at time $k$ due to Bob and Eve respectively and $k=1,\ldots,N$. Fig. \ref{mse_time} shows the error vectors $\mathbf{e}_{AB}$ (MSE Bob) and $\mathbf{e}_{AE}$ (MSE Eve) with Bob and Eve as transmitters located at the same place and Alice as the receiver exemplary. The distance between Alice and Bob and Alice and Eve is \mbox{$d_{AB}=d_{AE}=3$} m in all setups. The MSE is within the same range for both transmitters and they can obviously not be distinguished from each other based on that information. The same counts for the Pearson Correlation Coefficient (PCC) which is shown additionally and also indicates high similarity for both considered cases. However, the setup with Bob and Eve at different locations (Fig. \ref{results_space}) yields a different result. Fig. \ref{mse_space} shows the set of reference measurements that is obtained in order to determine the decision threshold $e_{th}$. Though Eve is located only $d_{BE}=10$ cm away from Bob in this setup, they can clearly be distinguished in case of MSE (Fig. \ref{mse_space}). In case of PCC (Fig. \ref{pcc_space}) there is still a huge amount of indistinguishable errors. With a measurement setup of $N=10000$ conducted measurements each time we get an average packet drop rate of $0 \%$ for received packets from the benign transmitter Bob in the setup with same transmitter locations of Bob and Eve and an average packet drop rate of $6.37 \%$ for received packets from the malicious transmitter Eve. This confirms the presumption made based on observation of Fig. \ref{results_time}. In the setup with the transmitters at different locations we get an average packet drop rate of $0 \%$ for Bob and an average packet drop rate of $97,28 \%$ for Alice transmitting. The decision threshold $e_{th}$ is as mentioned chosen from the reference measurement set shown in Fig. \ref{mse_space} and used for all measurement sets. It is calculated as
\begin{equation}
e_{th}=\frac{1}{2} \left|\frac{1}{N}\sum_{k=1}^N e_{AB_{ref}}^{(k)}-\frac{1}{N}\sum_{k=1}^N e_{AE_{ref}}^{(k)}\right|
\end{equation}
with $\mathbf{e}_{AB_{ref}}$ and $\mathbf{e}_{AE_{ref}}$ being the error vectors of the reference measurement set.

%

\section{Conclusion and Future Work}
\label{CONC}

We have shown a promising approach on how to secure MC-MTC by means of PHYSEC, actually Channel Profile Monitoring (CPM), in a very efficient way. The combination of both, MC-MTC and CPM is essential considering system efficiency, as both rely on frequent transmission of data packets and with this also channel estimation. We can reuse channel estimations and make decisions on integrity and authenticity of received data packets with little to no further effort.
For a more reliable evaluation of our setup, the application of further evaluation approaches shown in section \ref{related work} is essential. Especially machine learning approaches seem interesting to us. The finding of a decision threshold has to be considered and optimized further as well. Especially the packet drop rate for Eve of $97.28 \%$ at a different location is still to low to be considered as reliable. However, for Bob the packet drop rate was always $0 \%$ in that case. Additionally we also want to focus on a mobile setup with little to moderate velocities.\\
We have assumed here that communicating nodes have already carried out initial authentication, which to achieve is another issue and needs to be considered as well in future work. 

\section*{Acknowledgment}
This work has been supported by the Federal Ministry of Education and Research of the Federal Republic of Germany (Foerderkennzeichen 16KIS0267, HiFlecs and KIS4ITS0001, Iuno). The authors alone are responsible for the content of the paper.



\bibliographystyle{IEEEtran}
\bibliography{references_v1}

\begin{thebibliography}{10}
\providecommand{\url}[1]{#1}
\csname url@samestyle\endcsname
\providecommand{\newblock}{\relax}
\providecommand{\bibinfo}[2]{#2}
\providecommand{\BIBentrySTDinterwordspacing}{\spaceskip=0pt\relax}
\providecommand{\BIBentryALTinterwordstretchfactor}{4}
\providecommand{\BIBentryALTinterwordspacing}{\spaceskip=\fontdimen2\font plus
\BIBentryALTinterwordstretchfactor\fontdimen3\font minus
  \fontdimen4\font\relax}
\providecommand{\BIBforeignlanguage}[2]{{%
\expandafter\ifx\csname l@#1\endcsname\relax
\typeout{** WARNING: IEEEtran.bst: No hyphenation pattern has been}%
\typeout{** loaded for the language `#1'. Using the pattern for}%
\typeout{** the default language instead.}%
\else
\language=\csname l@#1\endcsname
\fi
#2}}
\providecommand{\BIBdecl}{\relax}
\BIBdecl

\bibitem{Guillaume.}
R.~Guillaume, F.~Winzer, A.~Czylwik, C.~T. Zenger, and C.~Paar, ``Bringing
  phy-based key generation into the field: An evaluation for practical
  scenarios,'' in \emph{IEEE Vehicular Technology Conference (VTC Fall)}, 2015.

\bibitem{Zenger.2014}
C.~T. Zenger, M.-J. Chur, J.-F. Posielek, C.~Paar, and G.~Wunder, ``A novel key
  generating architecture for wireless low-resource devices,'' in
  \emph{In­ter­na­tio­nal Work­shop on Se­cu­re In­ter­net of Things
  (SIoT)}, 2014.

\bibitem{Ambekar.2014}
A.~Ambekar and H.~D. Schotten, ``Enhancing channel reciprocity for effective
  key management in wireless ad-hoc networks,'' in \emph{IEEE Vehicular
  Technology Conference (VTC Spring)}, 2014.

\bibitem{Xiao.2007}
{L. Xiao, L. Greenstein, N. Mandayam and W. Trappe}, ``Fingerprints in the
  ether: Using the physical layer for wireless authentication,'' in \emph{IEEE
  International Conference on Communications (ICC)}, 2007.

\bibitem{Xiao.2008}
L.~Xiao, L.~Greenstein, N.~Mandayam, and W.~Trappe, ``Using the physical layer
  for wireless authentication in time-variant channels,'' \emph{IEEE
  Transactions on Wireless Communications}, vol.~7, no.~7, pp. 2571--2579,
  2008.

\bibitem{Pei.2014}
C.~Pei, N.~Zhang, X.~S. Shen, and J.~W. Mark, ``Channel-based physical layer
  authentication,'' in \emph{IEEE Global Communications Conference (GLOBECOM)},
  2014.

\bibitem{Gulati.2013}
N.~Gulati, R.~Greenstadt, K.~R. Dandekar, and J.~M. Walsh, ``Gmm based
  semi-supervised learning for channel-based authentication scheme,'' in
  \emph{IEEE Vehicular Technology Conference (VTC Fall)}, 2013.

\bibitem{Tugnait.2010}
J.~K. Tugnait and H.~Kim, ``A channel-based hypothesis testing approach to
  enhance user authentication in wireless networks,'' in \emph{International
  Conference on COMmunication Systems and NETworks (COMSNETS 2010)}, 2010.

\bibitem{Shi.2013}
L.~Shi, M.~Li, S.~Yu, and J.~Yuan, ``Bana: Body area network authentication
  exploiting channel characteristics,'' \emph{IEEE Journal on Selected Areas in
  Communications}, vol.~31, no.~9, pp. 1803--1816, 2013.

\bibitem{Refaey.2014}
A.~Refaey, W.~Hou, and K.~Loukhaoukha, ``Multilayer authentication for
  communication systems based on physical-layer attributes,'' \emph{Journal of
  Computer and Communications}, vol.~2, no.~8, pp. 64--75, 2014.

\bibitem{Osman.2015}
N.~C.~Y. Osman, Y.-P.~E. Wang, N.~A. Johansson, N.~Brahmi, S.~A. Ashraf, and
  J.~Sachs, ``Analysis of ultra-reliable and low-latency 5g communication for a
  factory automation use case,'' in \emph{IEEE International Conference on
  Communication workshop}, 2015.

\bibitem{Schmidl.1997}
T.~M. Schmidl and D.~C. Cox, ``Robust frequency and timing synchronization for
  ofdm,'' \emph{IEEE Transactions on Communications}, vol.~45, no.~12, pp.
  1613--1621, 1997.

\end{thebibliography}
%

%
%
%

\end{document}